\documentclass[final,5p,times,twocolumn]{elsarticle}

\usepackage{amssymb}
\usepackage{amsmath}

\usepackage[colorlinks=true, linkcolor=blue, citecolor=blue, urlcolor=blue]{hyperref}

\usepackage{xcolor}
\journal{Physics Letters A}

\begin{document}



\begin{frontmatter}
\title{Quantum Energy Teleportation on Mixed Steady States Prepared by Equilibrium and Nonequilibrium Environments}


\author[addr1,addr2]{Xiaokun Yan}

\author[addr3,addr4,addr5]{Kun Zhang}
\ead{kunzhang@nwu.edu.cn} 

\author[addr6]{Jin Wang}
\ead{jin.wang.1@stonybrook.edu} 

\address[addr1]{College of Physics, Jilin University, Changchun 130022, China}
\address[addr2]{State Key Laboratory of Electroanalytical Chemistry, Changchun Institute of Applied Chemistry, Chinese Academy of Sciences, Changchun 130022, China}
\address[addr3]{School of Physics, Northwest University, Xi’an 710127, China}
\address[addr4]{Shaanxi Key Laboratory for Theoretical Physics Frontiers, Xi'an 710127, China}
\address[addr5]{Peng Huanwu Center for Fundamental Theory, Xi'an 710127, China}
\address[addr6]{Department of Chemistry, Stony Brook University, and Department of Physics and Astronomy, Stony Brook University, Stony Brook, New York 11794, USA}

\begin{abstract}
Quantum energy teleportation (QET), implemented via local operations and classical communication, enables carrier-free energy transfer by exploiting quantum resources. Although QET has been extensively studied theoretically and demonstrated experimentally in various quantum platforms, enhancing its performance for mixed states under environmental coupling remains a significant challenge. In this work, we investigate QET on mixed steady states of a two-qubit system coupled to equilibrium with detailed balence or nonequilibrium without detailed balence environments. Using the Bloch-Redfield master equation, we systematically examine how temperature difference and chemical potential difference as the source of nonequilibrium or degree of detailed balance breaking on the teleported energy. We show that the teleported energy of a mixed steady state is often governed by the eigenstate with the largest population, and that nonequilibrium environments can enhance the teleported energy in certain parameter regimes.
\end{abstract}


\end{frontmatter}




\section{Introduction}

Quantum teleportation (QT) is a well-known protocol that transfers an unknown quantum state to a distant location by using quantum entanglement together with local operations and classical communication (LOCC) \cite{bennett1993teleporting,bouwmeester1997experimental,boschi1998experimental}. Subsequently, Hotta introduced a novel protocol called quantum energy teleportation (QET), which enables the extraction of energy from the ground state via entanglement and LOCC \cite{hotta2008quantum}. QET has been theoretically investigated in a variety of physical systems, including spin chains \cite{hotta2008protocol,hotta2009quantum,hotta2010energy}, cold trapped ions \cite{hotta2009quantum}, harmonic chains \cite{nambu2010quantum}, and quantum fields \cite{hotta2008quantum,hotta2010quantum,hotta2010controlled,funai2017engineering,ikeda2023criticality}. It has also been studied in holographic conformal field theory \cite{giataganas2016towards}. Recently, QET was experimentally demonstrated in the laboratory and on a quantum chip \cite{rodriguez2023experimental,ikeda2023demonstration}. Although QT and QET both rely on LOCC, their goals differ: QT transmits quantum state information, whereas QET aims to extract energy from a local subsystem rather than to reconstruct the state.

In the original QET protocol \cite{hotta2008quantum}, the sender (Alice) and the receiver (Bob) share the ground state, which is a strong local passive (SLP) state \cite{frey2014strong,PhysRevLett.123.190601}. Therefore, Bob cannot extract energy from his subsystem by any local operation alone. QET enables energy extraction from a distant subsystem by exploiting two key quantum features of the ground state: zero-point fluctuations and entanglement. In the ground state, local subsystems exhibit quantum fluctuations with nonvanishing energy density, and these fluctuations are spatially entangled across distant regions. The local measurement performed by Alice injects energy into her subsystem and simultaneously provides information about distant fluctuations through entanglement. This measurement outcome is then communicated classically to Bob, who performs a conditional local operation designed to suppress specific fluctuation modes in his region. As a result, negative local energy density can emerge at Bob's site, allowing positive energy to be extracted. Thus, energy is effectively ``teleported'' not through the transmission of energetic carriers, but through LOCC-assisted control enabled by ground-state entanglement. In this sense, Alice's measurement does not directly transmit energy to Bob; rather, it provides information that allows Bob to extract energy locally. It has been shown that, in the QET protocol, the energy extracted by Bob is smaller than the energy injected by Alice \cite{hotta2008protocol}, and therefore energy conservation is not violated.

Several generalized QET protocols have been proposed in recent years. For example, Wu et al. identified SLP states in the minimal QET model and showed that QET can extract energy from such states despite the impossibility of direct local extraction \cite{PhysRevA.110.052424}. They also suggested that this approach may be extended to mixed states, although such an extension has not been established in full generality. Furthermore, Kazuki Ikeda proposed extending the concept of QET beyond energy to arbitrary observables \cite{ikeda2025beyond}. To illustrate this idea, he studied a (1+1)-dimensional Dirac system and used feedback control based on fermion chirality to activate electric current and charge, deriving a rigorous upper bound on the teleported quantity. In conventional QET, the upper bound on energy output is severely constrained by distance; however, this limitation can be overcome by employing squeezed vacuum states with local vacuum regions between the two parties \cite{hotta2014quantum}. In addition, long-range QET can be realized in a hyperbolic quantum network by transmitting local quantum information via quantum teleportation and performing conditional operations on that information \cite{ikeda2024long}.

QET necessarily requires quantum resources, but which resource is operationally relevant depends on the specific setting, and no universal consensus has yet been reached. In the minimal QET model, Lin et al. found that initial-state entanglement and coherence show no clear relationship with the extractable energy, although they correlate positively with the energy-output efficiency \cite{fan2024role}. Moreover, the change in system entropy during the measurement process sets a lower bound on the transferable energy \cite{hotta2010energy}. For thermal states, QET is enabled by thermal discord \cite{hotta2013quantum}. However, in some cases, quantum discord is not the relevant resource for QET, as shown for a three-spin Ising chain in a Gibbs state \cite{trevison2015quantum}. More generally, the total amount of transmitted energy and information is constrained by entanglement \cite{wang2024quantum}.

In practice, quantum systems inevitably interact with their environments. Since Alice and Bob are spatially separated, it is essential to account for the effects of their distinct local environments. In the standard QET model, the system is assumed to be in the ground state, which typically yields both low transferred energy and low efficiency \cite{hassan2024enhanced}. However, when the system is prepared in a mixed state, the presence of excited-state populations need not be detrimental to energy extraction. At the same time, for mixed states it becomes essential to distinguish genuine QET-assisted extraction from energy that Bob could already extract by a local unitary (LU) operation alone. In the QET protocol considered here, Bob's operation is restricted to a LU. In the Appendix, we therefore evaluate the amount of energy extractable by LU-only operations and compare it with the energy output of QET. This comparison shows that QET not only enables energy extraction from local passive (LP) states, from which no energy can be extracted by LU operations alone, but also yields a larger extracted energy than LU-only operations even when the state is not LP.

Here we consider QET in a two-qubit model where each qubit interacts only with its own environment, and we investigate how equilibrium and nonequilibrium reservoirs can be exploited to improve QET performance. Notably, previous studies have shown that nonequilibrium environments can enhance various quantum correlations, including entanglement \cite{yu2004finite,carvalho2004decoherence,harlender2022transfer,wang2018steady}, quantum discord \cite{mazzola2010sudden,li2021quantum,radhakrishnan2020multipartite}, quantum steering \cite{zhang2021asymmetric}, Bell nonlocality \cite{zhang2021entanglement}, and temporal correlations \cite{castillo2013enhanced,zhang2020influence}. Nonequilibrium steady states exhibit properties distinct from equilibrium states \cite{wu2011quantum,lambert2007nonequilibrium,quiroga2007nonequilibrium,sinaysky2008dynamics}. We adopt the Bloch--Redfield master equation to describe the nonequilibrium two-qubit model, which allows us to analyze changes in energy output induced by temperature or chemical-potential differences between the baths \cite{PhysRev.105.1206,redfield1957theory,redfield1996relaxation,ishizaki2009adequacy,lee2015coherent,novoderezhkin2004coherent,jeske2015bloch}. Compared with secular Lindblad-type treatments, the Bloch--Redfield equation without the secular approximation retains additional couplings between populations and coherences and is therefore better suited to describing biased steady states in the present setting \cite{zhang2014curl,li2015steady,huangfu2018steady,zhang2015landscape,zhang2015shape,wang2018coherence,guarnieri2018steady}. Its known limitations regarding density-matrix positivity, as well as standard strategies for mitigating such issues, have been discussed extensively in the literature \cite{ishizaki2009adequacy,jeske2015bloch,spohn1980kinetic,suarez1992memory}. Additionally, we consider the effect of detuning between the qubit energy levels, which can further enhance nonequilibrium effects.

We find that a temperature difference in bosonic reservoirs consistently suppresses the QET energy output, whereas in fermionic reservoirs it can enhance the energy output. The chemical-potential difference also has a pronounced impact: when the average chemical potential is far below or far above the system energy levels, the energy output is reduced; by contrast, when the chemical potential is comparable to the system energy levels, the energy output can be enhanced over a finite parameter range. For steady states dominated by low-excitation populations, increasing Alice's energy level can improve the energy output, whereas for high-excitation steady states, increasing Bob's energy level can likewise enhance the energy output.

The paper is organized as follows. In Sec. \ref{II} we introduce the standard QET protocol and analyze the energy output when QET is performed on each eigenstate of the Hamiltonian. We also review the Redfield master equation used in our study. QET under equilibrium and nonequilibrium environments is analyzed in Secs. \ref{III} and \ref{IIII}, respectively. Finally, in Sec. \ref{V} we summarize our findings. For simplicity, we set $\hbar = k_B= 1$ in the following sections.

\section{Energy Teleportation and Redfield Equation}\label{II}

In this section, we first review the QET protocol in Sec.~\ref{II_A}. We then analyze the energy output for an initial mixed state with X structure in Sec.~\ref{II_B}. Finally, in Sec.~\ref{II_C}, we introduce the model used in this work, namely two qubits coupled to nonequilibrium environments.

\subsection{Two-qubit Model of Energy Teleportation}\label{II_A}

The minimal QET model, also known as the two-particle Hotta model \cite{hotta2011quantum}, considers two interacting spin-$1/2$ particles as qubits $A$ and $B$, possessed by Alice and Bob, respectively. In the original formulation, the Hamiltonian is chosen such that the ground-state energy is zero, and the protocol is performed on the ground state. In more general QET settings, however, the initial state may be mixed \cite{PhysRevA.110.052424}. The system Hamiltonian is given by
\begin{align}\label{HS}
    H_{AB}=H_A+H_B+V=\varepsilon_A \sigma^z_A + \varepsilon_B \sigma^z_B + 2 \kappa \sigma^x_A\sigma^x_B,
\end{align}
where $\varepsilon_{A,B}$ are the qubit energy splittings, $\kappa$ is the interaction strength between qubits $A$ and $B$, and $\sigma^z_{A,B}$ and $\sigma^x_{A,B}$ are the Pauli operators acting on qubits $A$ and $B$, respectively.

In the original QET model, the energy levels satisfy $\varepsilon_A=\varepsilon_B$. In the present work, we relax this constraint and consider detuned qubit energy levels, so that the system becomes asymmetric and the ground-state energy is not necessarily set to zero \cite{PhysRevA.110.052424}. The eigenvalues of the Hamiltonian in Eq.~(\ref{HS}) are
\begin{align}
   E_1&=-\sqrt{\Omega^2+4\kappa^2},\nonumber \\
   E_2&=-\sqrt{\Delta^2+4\kappa^2}, \nonumber\\
   E_3&=\sqrt{\Delta^2+4\kappa^2},\nonumber \\
   E_4&=\sqrt{\Omega^2+4\kappa^2},
\end{align}
and the corresponding eigenstates are
\begin{align}
\label{eq:eigenstates}
    |E_1\rangle &=-\sin\phi_1|11\rangle+\cos\phi_1|00\rangle,\nonumber \\
    |E_2\rangle &=-\sin\phi_2|10\rangle+\cos\phi_2|01\rangle,\nonumber \\
    |E_3\rangle &=\cos\phi_2|10\rangle+\sin\phi_2|01\rangle,\nonumber \\
    |E_4\rangle &=\cos\phi_1|11\rangle+\sin\phi_1|00\rangle,
\end{align}
where $\Omega=\varepsilon_A+\varepsilon_B$ and $\Delta=\varepsilon_A-\varepsilon_B$. The angles $\phi_1$ and $\phi_2$ are given by
\begin{align}
    \phi_1 & =\arctan\left(\frac{2\kappa}{\Omega+\sqrt{\Omega^2+4\kappa^2}}\right), \nonumber \\
    \phi_2 & =\arctan\left(\frac{2\kappa}{\Delta+\sqrt{\Delta^2+4\kappa^2}}\right).
\end{align}

The QET protocol consists of three steps \cite{hotta2011quantum}: (i) Alice first performs the projective measurement
\begin{equation}
    P_A(u)=\frac{1}{2}(I+u\sigma^x_A),
\end{equation}
on qubit $A$ and obtains the outcome $u\in\{\pm1\}$; (ii) Alice then communicates the measurement outcome $u$ to Bob through a classical channel. To ensure that the energy injected by Alice during the measurement does not propagate to Bob before his local operation, the time required for the classical communication and Bob's operation must be much shorter than the timescale $1/(2\kappa)$ associated with energy propagation through the interaction \cite{rodriguez2023experimental,ikeda2023demonstration}. This separation of timescales is a standard requirement of the protocol and is experimentally achievable; (iii) Bob performs a LU operation $U_B(u)$ conditioned on the value of $u$. The operation $U_B(u)$ is given by
\begin{align}
    U_B(u)=I\cos\theta-iu\sigma^y_B\sin\theta,
\end{align}
where $\theta$ is a real control parameter.

For a mixed initial state $\rho_{AB}$, the energy is
\begin{align}
    E_0(\rho_{AB})=\text{Tr}(H_{AB}\rho_{AB}).
\end{align}
After Alice performs the projective measurement $P_A(u)$, the average energy of the system is given by
\begin{align}
    E_A(\rho_{AB})=\sum_{u=\pm1} \text{Tr}\left(H_{AB}P_A(u)\rho_{AB}P_A^\dagger(u)\right).
\end{align}
The measurement $P_A(u)$ affects only the energy of subsystem $A$, while the energy of subsystem $B$ remains unchanged, because
\begin{align}
    [P_A(u),H_B]=[P_A(u),V]=0.
\end{align}
After Alice communicates the measurement outcome $u$ to Bob, Bob applies $U_B(u)$ to his qubit. The energy of the system then becomes
\begin{align}
    E_B(\rho_{AB})=\sum_{u=\pm1} \text{Tr}\left(H_{AB}U_B(u)P_A(u)\rho_{AB}P_A^\dagger(u)U_B^\dagger(u)\right).
\end{align}
The energy difference $E_{\text{out}}=E_A-E_B$ represents the energy output extracted at Bob's side with the assistance of Alice's measurement and classical communication.

\begin{figure}[t!]
    \centering
    \includegraphics[width=0.48\textwidth]{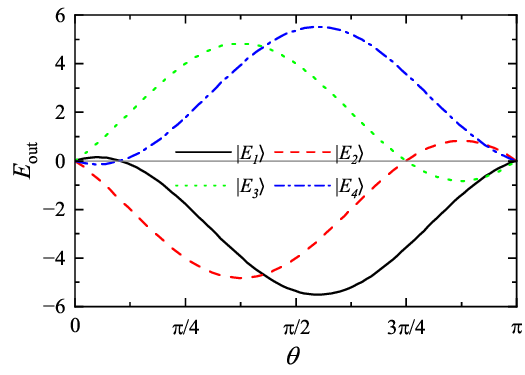}
    \caption{Energy output of four eigenstates of $H_{AB}$ (\ref{HS}). The parameters are set as $\kappa=1$ and $\varepsilon_A=\varepsilon_B=2$.}
    \label{eigf}
\end{figure}

Suppose that the mixed initial state $\rho_{AB}$ is a classical mixture of the four eigenstates, for example, a thermal state. Before analyzing QET for the mixture $\rho_{AB}$, we first calculate the energy output $E_{\text{out}}$ for the individual eigenstates:
\begin{multline}
    E_{\text{out}}(|E_1\rangle)=-E_{\text{out}}(|E_4\rangle) \\  =\frac{1}{\sqrt{\Omega^2+4\kappa^2}}\left(2 \varepsilon_A \kappa \sin2\theta - (\varepsilon_B\Omega+4\kappa^2)(1-\cos2\theta)\right), \nonumber 
\end{multline}
\begin{multline}
\label{eeig}
    E_{\text{out}}(|E_2\rangle)=-E_{\text{out}}(|E_3\rangle)\\
    =\frac{1}{\sqrt{\Delta^2+4\kappa^2}}\left(-2 \varepsilon_A \kappa \sin2\theta + (\varepsilon_B\Delta-4\kappa^2)(1-\cos2\theta)\right).
\end{multline}
Clearly, the energy output depends on the parameter $\theta$ in the conditional unitary operation $U_B(u)$. However, there is no single optimal value of $\theta$ that maximizes all four values of $E_{\text{out}}$ simultaneously, as illustrated in Fig.~\ref{eigf}. This observation already suggests that, for mixed states, the optimal QET protocol depends nontrivially on the population distribution among the eigenstates.

It is evident from Eq.~(\ref{eeig}) and Fig.~\ref{eigf} that $E_{\text{out}}(|E_1\rangle)$ and $E_{\text{out}}(|E_4\rangle)$ exhibit opposite behavior, and so do $E_{\text{out}}(|E_2\rangle)$ and $E_{\text{out}}(|E_3\rangle)$. This implies that, when the parameter $\theta$ is chosen to maximize $E_{\text{out}}(|E_1\rangle)$ or $E_{\text{out}}(|E_2\rangle)$, the corresponding value of $E_{\text{out}}(|E_4\rangle)$ or $E_{\text{out}}(|E_3\rangle)$ is minimized. Therefore, for mixed states, the maximum energy output is determined by the density matrix resulting from the mixture of the four eigenstates. However, when one particular eigenstate dominates the mixture, the behavior of $E_{\text{out}}$ closely follows that of the dominant eigenstate. This observation enables a qualitative analysis of $E_{\text{out}}$ under specific conditions.

\subsection{Energy teleportation with $X$ state}\label{II_B}

The energy output in QET originates from the correlations between subsystems $A$ and $B$. However, the specific quantum resources underpinning energy teleportation have not yet been fully clarified. For the ground state, the efficiency of energy transfer is closely related to coherence and concurrence \cite{fan2024role}. However, for mixed states, it remains unclear which quantum resources are primarily responsible for the energy output. Although the presence of quantum resources can enhance the energy output, neither the extracted energy nor the transfer efficiency necessarily varies monotonically with any single quantum resource. In more extreme cases, energy can be extracted even when QET is performed on a direct product state \cite{haque2024aspects}. Although no one-to-one correspondence between the energy output and a specific quantum resource has been established, we can still analyze the energy output from the structure of the initial state. 

Consider the mixed state $\rho_{AB}$ in our model with the 'X'-structure from the Hamiltonian in Eq. (\ref{HS}), as expressed in the form 
\begin{equation}
    \rho_{AB}^{X}=\begin{pmatrix}
  a & 0 & 0 & \chi e^{i\nu }\\
 0 & b & \delta  e^{i\epsilon  } & 0 \\
  0& \delta  e^{-i\epsilon  } & c & 0\\
  \chi e^{-i\nu }& 0 & 0 & d
\end{pmatrix}\label{rhoX},
\end{equation}
The energy output $E_{\text{out}} = E_A-E_B$ is given by
\begin{align}
    E_{\text{out}}\left(\rho_{AB}^{X}\right)=D\sin2 \theta-F(1- \cos2 \theta),
\end{align}
where 
\begin{align}
    D=2 (-a + b - c + d) \kappa + \varepsilon_B (\delta \cos\epsilon +  \chi \cos\nu),\nonumber\\
    F=-(a - b + c - d) \varepsilon_B - 4\kappa (\delta \cos\epsilon +  \chi \cos\nu).\nonumber
\end{align}
It is evident that the output of energy is dependent on the parameter $\theta$. The maximal value is given by
\begin{align}
\label{eq:optimal_theta}
    \tan(2\theta_1)=\frac{D}{F}\quad\text{or}\quad\tan\left(2\theta_2+\frac{\pi}{2}\right)=\frac{D}{F},
\end{align}
and the corresponding energy output is
\begin{align}
\label{eq:E_max}
E^\text{max}_{\text{out}}\left(\rho_{AB}^{X}\right)=\sqrt{D^2+F^2}-F.
\end{align}
Note that the optimal value of $\theta_{1,2}$ is not unique; however, the maximal value of $E_{\text{out}}$ is the same in both cases.

\subsection{Environments and Bloch-Redfield equation}\label{II_C}

The mixed initial state $\rho_{AB}$ is prepared by the environments. We consider a scenario in which each qubit is coupled to its own reservoir, with the two reservoirs allowed to have different temperatures or chemical potentials. In this setting, the environments are treated primarily as state-preparation and reset mechanisms, rather than as disposable components of the protocol. Accordingly, we take the steady state of the open system as the initial state for QET. After one round of the protocol, the environments drive the system back to the same steady state, thereby enabling repeated operation.

The total Hamiltonian combining the system and the environment is given by
\begin{equation}
    H = H_{AB} + H_R + H_I,
\end{equation}
where $ H_{AB} $ is the Hamiltonian of the two interacting qubits, as defined in Eq. (\ref{HS}). The free Hamiltonian of the reservoirs, $ H_R $, is
\begin{align}
    H_R = \sum_{k_A} \omega_{k_A} b_{k_A}^\dagger b_{k_A} + \sum_{k_B} \omega_{k_B} b_{k_B}^\dagger b_{k_B},
\end{align}
where $ b_{k_A} $ ($ b_{k_A}^\dagger $) and $ b_{k_B} $ ($ b_{k_B}^\dagger $) are the annihilation (creation) operators for the $ k $-th mode with frequencies $ \omega_{k_A} $ and $ \omega_{k_B} $ of the reservoirs coupled to qubits $ A $ and $ B $, respectively. The qubit-reservoir interaction under the rotating wave approximation is
\begin{multline}
    H_I = \sum_{k_A} g_{k_{A}}\left(\sigma^{-}_{A} b_{k_{A}}^{\dagger}+\sigma^{+}_{A} b_{k_{A}}\right) 
     +\sum_{k_B} g_{k_{B}}\left(\sigma^{-}_{B} b_{k_{B}}^{\dagger}+\sigma^{+}_{B} b_{k_{B}}\right),
\end{multline} 
where $g_{k_{A}}$ and $g_{k_{B}}$ are qubit-reservoir coupling strengths. In the eigenbasis of $H_{S}$ (\ref{HS}), interaction Hamiltonian $H_I$ can be rewritten as
\begin{multline}
    H_I=\sum_{k_A} g_{k_{A}}(\eta_A+\xi_A) b_{k_{A}}^{\dagger} 
     + \sum_{k_B}g_{k_{B}}(\eta_B+\xi_B) b_{k_{B}}^{\dagger} +\text{H.c.},
\end{multline}
where $\eta_{A,B}, \xi_{A,B}$ are transition operators given by 
\begin{align}
 \eta _ { A } & = \sin (\phi_1+\phi_2) ( | E_3 \rangle \langle E_4 | - | E_1 \rangle \langle E_2 | ) , \nonumber\\ \eta _ { B } & = \cos (\phi_1-\phi_2) ( | E_3 \rangle \langle E_4 | + | E_1 \rangle \langle E_2 | ) , \nonumber\\ \xi _ { A } & = \cos (\phi_1+\phi_2)( | E_2 \rangle \langle E_4 | + | E_1 \rangle \langle E_3 | ) , \nonumber\\ \xi _ { B } & = \sin (\phi_1-\phi_2) ( | E_2 \rangle \langle E_4 | - | E_1 \rangle \langle E_3 | ).
\end{align}
The corresponding transition frequencies are 
\begin{equation}
    \varepsilon_{\pm}=\sqrt{\Omega^2+4\kappa^2}\pm \sqrt{\Delta^2+4\kappa^2}.
\end{equation}
Here $\varepsilon_{-}$ corresponds to the transitions from the state $| E_2 \rangle$ to $| E_1 \rangle$ and the state $| E_4 \rangle$ to $| E_3 \rangle$. The transition frequency $\varepsilon_{+}$ corresponds to the transitions from the state $| E_4 \rangle$ to $| E_2 \rangle$ and the state $| E_3 \rangle$ to $| E_1 \rangle$. 


The Born-Markov quantum master equation in the interaction picture is given by \cite{PhysRev.105.1206,redfield1957theory}
\begin{equation}
    \frac{d \tilde{\rho}_{AB}}{dt}=-\int_{0}^\infty ds~ \text{Tr}_R\left[\tilde H_I(t),[\tilde H_I(t-s),\tilde{\rho}_{AB}\otimes\tilde{\rho}_R]\right],
\end{equation}
where $\tilde{\rho}_{AB}$ is the reduced density operator of the coupled two qubits in the interaction picture, and $\tilde{\rho}_{R}$ is the initial state of the reservoirs, assuming in its own equilibrium state. Going back to the Schr$\ddot{\mathrm{o}}$dinger picture, the Bloch-Redfield equation is given by
\begin{equation}
    \frac{d\rho_{AB}}{dt}=-i[H_{AB},\rho_{AB}]+\sum_{j=A,B} \mathcal { D }_j(\rho_{AB}), 
\end{equation}
where $\mathcal {D}_j(\rho_{AB})$ is the dissipator given by
\begin{multline}
    \mathcal { D } _ { j } (\rho_{AB}) = \\
     \alpha _ { j } ( \varepsilon _ { - } ) ( \eta _ { j } ^ { \dagger } \rho_{AB} \eta _ { j } + \eta _ { j } ^ { \dagger } \rho_{AB} \xi _ { j } - \eta _ { j } \eta _ { j } ^ { \dagger } \rho_{AB} - \xi _ { j } \eta _ { j } ^ { \dagger } \rho_{AB})  \\
    + \alpha _ { j } ( \varepsilon _ { + } ) ( \xi _ { j } ^ { \dagger } \rho_{AB} \xi _ { j }   + \eta _ { j }^ { \dagger }   \rho_{AB} \xi _ { j }- \xi _ { j }\xi _ { j } ^ { \dagger }  \rho_{AB} - \eta _ { j }\xi _ { j } ^ { \dagger }  \rho_{AB}) \\
    + \beta _ { j } ( \varepsilon _ { - } )  ( \eta _ { j }  \rho_{AB} \eta _ { j }^ { \dagger } + \eta _ { j }  \rho_{AB} \xi _ { j }^ { \dagger } - \eta _ { j } ^ { \dagger }\eta _ { j }  \rho_{AB} - \xi _ { j } ^ { \dagger }  \eta _ { j }\rho_{AB})  \\
    + \beta _ { j } ( \varepsilon _ { +} )( \xi _ { j }  \rho_{AB} \xi _ { j }^ { \dagger }   + \eta _ { j }   \rho_{AB} \xi _ { j }^ { \dagger }- \xi _ { j }^ { \dagger } \xi _ { j }  \rho_{AB} - \eta _ { j }^ { \dagger } \xi _ { j }  \rho_{AB}) \\
    + \text{H.c.}
\end{multline}
Here the coefficients $\alpha _ { j }(\varepsilon)$ and  $\beta_ { j }(\varepsilon)$ are the dissipation rates, given by
\begin{align}
    \alpha _ { j }(\varepsilon)= & \gamma_j(\varepsilon)n_j(\varepsilon), \nonumber \\
    \beta_ { j }(\varepsilon) = & \gamma_j(\varepsilon)(1\pm n_j(\varepsilon)),
\end{align}
where the coupling spectrum $\gamma_j(\varepsilon)$ is
\begin{align}
    \gamma_j(\varepsilon)=\pi \sum_{k_j}|g_{k_j}|^2\delta(\varepsilon-\omega_{k_j})
\end{align}
and $n_j(\varepsilon)$ is the Bose-Einstein (minus sign) or the Fermi-Dirac (plus sign) distribution
\begin{align}
    n_j(\varepsilon)=\frac{1}{e^{(\varepsilon-\mu_j)/T_j}\mp1}.
\end{align}
The sign of $ \beta_ { j }(\varepsilon)$ is plus for the bosonic reservoirs, while it is minus for fermionic reservoirs. Parameters $T_j$ and $\mu_j$ are the equilibrium temperatures and chemical potentials of $j$-th reservoir, respectively. For bosonic reservoirs, such as photon or phonon baths, the particle number is not conserved with a vanishing chemical potential. Since the Bloch-Redfield equation is based on the assumption that the interaction between the system and the environment is weak, we can further assume that the coupling spectra with different frequencies are much less than the energy scale of the two qubits, namely $g_{k_j} \ll \varepsilon_A,\varepsilon_B$. Therefore, it is reasonable to view $g_{k_j}$ as constants (independent of the transition frequencies $\varepsilon_{\pm}$), we set $g_{k_A}=g_A,g_{k_B}=g_B$. 

We consider QET under steady-state conditions, where, after completing a total protocol, the environment can “cool” the system back to its initial state. In the minimal QET model, one usually assumes \(\kappa \sim \varepsilon_{A,B}\) to enhance the performance of QET. Meanwhile, the execution time of the QET protocol must be shorter than the timescale of internal energy diffusion in the system, namely \(1/\kappa\). In the weak-coupling regime, one has \(\gamma_{A,B}\sim g_{A,B}^2 \ll \varepsilon_{A,B} \sim \kappa\), which implies that the timescale of protocol execution $1/\kappa$ is much shorter than the timescale on which the environment affects the system $1/\gamma_{A,B}$. Therefore, the environmental influence can be neglected during the execution of the protocol.

The two-qubit steady state can be obtained by reformulating the Bloch--Redfield equation in Liouville space \cite{wang2018steady,wang2019nonequilibrium}. It corresponds to the right eigenvector of the superoperator with zero eigenvalue. Because the eigenstates of $H_{AB}$ lead to an $X$-state structure [cf. Eq.~(\ref{rhoX})], the steady state retains an $X$-type density-matrix form, both in the Hamiltonian eigenbasis and in the local basis. For all steady states presented in this work, we numerically verified positivity and unit trace before evaluating the QET energy output. In other words, for each parameter set shown in the figures, the steady-state density matrix satisfies \(\rho_{\mathrm{AB}}\ge 0\) and \(\mathrm{Tr}(\rho_{\mathrm{AB}})=1\) within numerical precision. Therefore, the results reported here are restricted to parameter regimes in which the Bloch--Redfield steady state remains physical.

In this work, we analyze QET mainly from the perspective of energy-eigenstate populations, rather than attempting to identify a specific entanglement resource. This perspective is motivated by the analytical expression for the energy output, where the environmental influence on the steady state is most directly reflected in the redistribution of the weights of different energy eigenstates. Therefore, the energy-eigenstate populations provide a natural bridge between the bath-induced steady state and the resulting QET performance, and their variation captures the dominant qualitative change of the energy output.
It should be stressed, however, that two different notions of off-diagonal elements are involved. In the energy eigenbasis, the nonsecular Redfield equation may in principle generate off-diagonal steady-state coherences between different energy eigenstates. We have checked these terms numerically and found that, in the parameter regimes considered in this work, they are much smaller than the corresponding diagonal populations. Thus, the energy-basis coherences do not dominate the QET behavior reported below, and the population-based description provides an adequate qualitative characterization.
This should not be confused with the off-diagonal elements in the local computational basis, in which the QET protocol is actually implemented. Since the eigenstates of \(H_{AB}\) are coherent superpositions of computational-basis states, a state that is nearly diagonal in the energy eigenbasis can still contain sizable off-diagonal elements in the computational basis. These computational-basis off-diagonal elements are fully retained in the evaluation of \(E_{\mathrm{out}}\). Nevertheless, they are basis-dependent and do not provide a simple universal descriptor of the environmental effect on QET. For this reason, we use the energy-eigenstate populations as the main qualitative indicator, while the quantitative calculation is always performed using the full density matrix.

To characterize the nonequilibrium nature of the steady state, we define the currents induced by each reservoir through the dissipator \(\mathcal{D}_j(\rho_{AB})\). For the \(j\)-th reservoir, the energy current flowing from the reservoir into the system is defined as \(J_j^{E}=\mathrm{Tr}[H_{AB}\mathcal{D}_j(\rho_{AB})]\). For bosonic reservoirs, since the particle number is not conserved and the chemical potential vanishes, the heat current is identical to the energy current, namely \(J_j^{Q}=J_j^{E}=\mathrm{Tr}[H_{AB}\mathcal{D}_j(\rho_{AB})]\). For fermionic reservoirs, besides the energy current, one should also distinguish the particle current, defined as \(J_j^{P}=\mathrm{Tr}[N\mathcal{D}_j(\rho_{AB})]\), where \(N\) is the particle-number operator of the system. Accordingly, the heat current from the \(j\)-th fermionic reservoir is given by \(J_j^{Q}=J_j^{E}-\mu_j J_j^{P}\). Here, a positive current means that the corresponding quantity flows from the reservoir into the system. In the steady state, the currents associated with the two reservoirs are opposite to each other. For fermionic reservoirs, the energy and particle currents satisfy
\(J_A^E+J_B^E=0\) and \(J_A^P+J_B^P=0\) in the steady state without chemical potential difference.
However, when the two reservoirs have
different chemical potentials, the heat currents do not generally satisfy
\(J_A^Q+J_B^Q=0\). Instead,$J_A^Q+J_B^Q=(\mu_B-\mu_A)J_A^P,$
which represents the contribution associated with chemical work. Only
when \(\mu_A=\mu_B\), or when the particle current vanishes, do the heat
currents become equal in magnitude and opposite in direction.


\section{Quantum energy teleportation under equilibrium environments}\label{III}

We separately discuss the influence of equilibrium bosonic and fermionic environments on QET in Secs.~\ref{sec:eq_bosonic} and \ref{sec:eq_fermionic}, respectively.

\subsection{Equilibrium bosonic environments}\label{sec:eq_bosonic}

\begin{figure}[t!]
    \centering
    \includegraphics[width=0.48\textwidth]{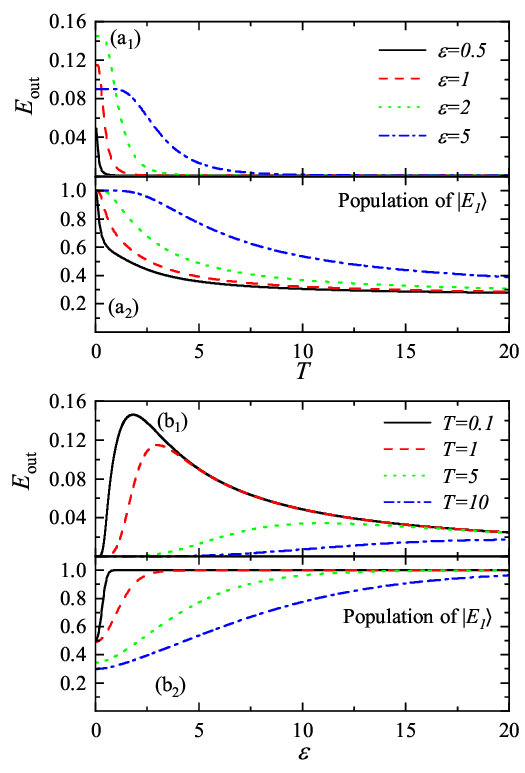}
    \caption{Energy output of steady states in the equilibrium bosonic reservoirs and the corresponding population of $|E_1\rangle$. ($\text{a}_1$) Energy output when the energy levels are set as $\varepsilon= 0.5~ (\text{black solid line})$, $1~(\text{red dashed line})$, $2~(\text{green dot line})$ and $5~(\text{blue dashed dot line})$. ($\text{a}_2$) The population of $|E_1\rangle$ corresponding to ($\text{a}_1$). ($\text{b}_1$) Energy output when the temperatures are set as $T= 0.1~ (\text{black solid line})$, $1~(\text{red dashed line})$, $5~(\text{green dot line})$ and $10~(\text{blue dashed dot line})$. ($\text{b}_2$) The population of $|E_1\rangle$ corresponding to ($\text{b}_2$). The other parameters are set as $\kappa=1$ and $g_A=g_B=0.05$.}
    \label{eqbte}
\end{figure}

Suppose that the initial state is a mixture of the eigenstates of $H_{AB}$, for example, a thermal state. Because the same control parameter $\theta$ must be used for the whole mixed state, the maximal energy output of the mixture is generally smaller than the weighted sum of the individually optimized outputs of the eigenstates. This suggests that the QET performance of mixed states is nontrivial. As shown in Eq.~(\ref{eeig}), we already have analytical expressions for the energy output of the four eigenstates. Therefore, the environmental influence on $E_{\text{out}}$ can be analyzed by studying how the populations of these eigenstates vary.

For bosonic reservoirs, the steady-state population distribution remains ordered: the ground state has the largest population, and in the high-temperature limit the populations of all eigenstates approach each other. As a result, the contribution of the excited states to the energy output is relatively limited. Although the maximal \(E_{\text{out}}\) of the excited states, when considered individually, can exceed that of the ground state, their contribution in the mixed state is not sufficient to overcome the dominant ground-state contribution under the same choice of $\theta$. Consequently, \(E_{\text{out}}\) is governed predominantly by the ground-state population.

In the bosonic case, when the two reservoirs are identical and share the same temperature, the system reaches the thermal equilibrium state associated with $H_{AB}$. We denote the equilibrium temperature by $T=T_A=T_B$. As shown in Fig.~\ref{eqbte} $(\text{a}_1)$, the energy output exhibits a short plateau as the temperature increases, followed by a rapid decline. The width of this plateau increases for systems with larger energy levels $\varepsilon$, which can be understood from the suppression of thermal excitation at low temperatures, as illustrated in Fig.~\ref{eqbte} $(\text{a}_2)$.

Furthermore, the energy output does not scale linearly with the energy level $\varepsilon$, as shown in Fig.~\ref{eqbte} $(\text{b}_1)$. When $\varepsilon=\varepsilon_A=\varepsilon_B$ increases at fixed equilibrium temperature, the population of $|E_1\rangle$ approaches unity, as shown in Fig.~\ref{eqbte} $(\text{b}_2)$. The energy output initially increases with $\varepsilon$, but decreases after the population of $|E_1\rangle$ becomes saturated. When the population of $|E_1\rangle$ approaches unity, the energy output reduces to
\[
E_\text{out}(|E_1\rangle)=\frac{4\kappa^2}{\sqrt{4\varepsilon^2 + 4\kappa^2}},
\]
as follows from Eq.~(\ref{eq:E_max}). Therefore, as $\varepsilon$ increases further, $E_{\text{out}}$ decreases.

\subsection{Equilibrium fermionic environments}

\label{sec:eq_fermionic}

\begin{figure}[t!]
    \centering
    \includegraphics[width=0.48\textwidth]{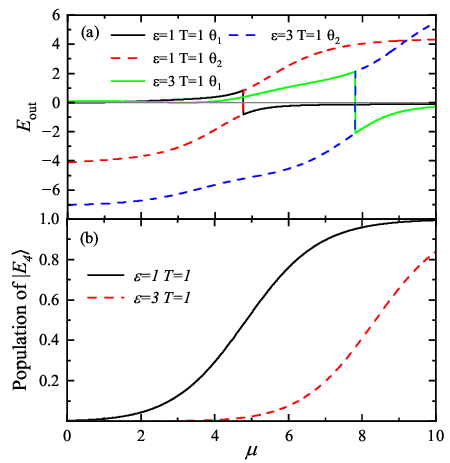}
    \caption{(a) Energy output of steady states with increasing chemical potential in fermionic reservoirs. The parameters are set as $\varepsilon=1$ and $\theta=\theta_1$ (black solid line), $\varepsilon=1$ and $\theta=\theta_2$ (red dashed line), $\varepsilon=3$ and $\theta=\theta_1$ (green solid line), $\varepsilon=3$ and $\theta=\theta_2$ (blue dashed line). (b) The population of state $|E_4\rangle$ with $\mu$. The parameters are set as $\varepsilon=1$ (black solid line) or $\varepsilon=3$ (red dashed line). The other parameters are set as $\kappa=1$, $T = T_A=T_B=1$, and $g_A=g_B=0.05$.}
    \label{eqfmu}
\end{figure}

Consider fermionic reservoirs with identical equilibrium temperatures and chemical potentials, $\mu=\mu_A=\mu_B$. When the equilibrium chemical potential exceeds the system energy levels, the population of the highest excited state $|E_4\rangle$ becomes dominant. In this regime, the QET protocol with $\theta=\theta_1$ given by Eq.~(\ref{eq:optimal_theta}) yields a negative energy output. By contrast, using the other optimal solution, $\theta=\theta_2$ [Eq.~(\ref{eq:optimal_theta})], significantly enhances the energy output without violating energy conservation, as shown in Fig.~\ref{eqfmu}.

We examine the dependence of the energy output on the equilibrium chemical potential $\mu$ for the two choices of $\theta$, as shown in Fig.~\ref{eqfmu} (a). As $\mu$ increases, population inversion develops among the eigenstates. At sufficiently high chemical potential, the excited states acquire significant populations. Figure~\ref{eqfmu} (b) shows that the population of the highest excited state $|E_4\rangle$ increases monotonically with $\mu$. When $\mu$ exceeds the system energy levels, the energy-output curves corresponding to $\theta_1$ and $\theta_2$ switch abruptly. This indicates that, once the population of $|E_4\rangle$ becomes dominant, the energy output of the mixed state approaches that of the pure state $|E_4\rangle$.

\section{Quantum energy teleportation under nonequilibrium environments}\label{IIII}

We separately discuss the influence of bosonic and fermionic nonequilibrium environments on QET in Secs. \ref{sec:noneq_bosonic} and \ref{sec:noneq_fermionic} respectively. 

\subsection{Nonequilibrium bosonic environments}\label{sec:noneq_bosonic}

\begin{figure*}[t!]
    \centering
    \includegraphics[width=\textwidth]{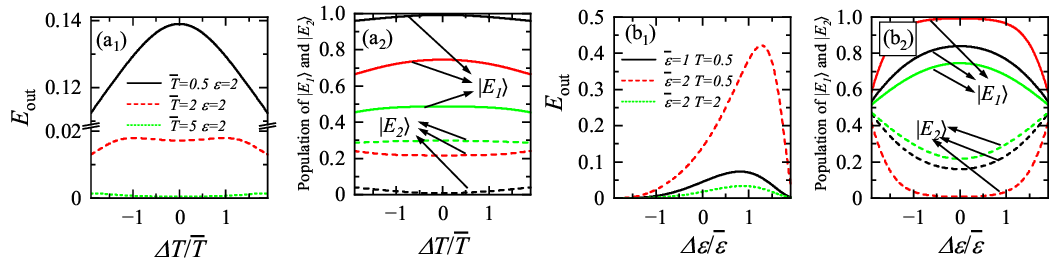}
    \caption{Energy output of steady states under the nonequilibrium bosonic environments or the energy detuning, and the corresponding population of eigenstates. ($\text{a}_1$) The average temperatures are set as $\bar{T}=0.5$ (black solid line), $\bar{T}=2$ (red dashed line), and $\bar{T}=5$ (green dot line). The energy levels are set as $\varepsilon_A=\varepsilon_B=2$. ($\text{a}_2$) The population of state $|E_1\rangle$ (in solid line) and state $|E_2\rangle$ (in dashed line) corresponding to ($\text{a}_1$). ($\text{b}_1$) The energy levels are set as $\bar{\varepsilon}=1$ with $T=0.5$ (black solid line), $\bar{\varepsilon}=2$ with $T=0.5$ (red dashed line), and $\bar{\varepsilon}=2$ with $T=2$ (green dot line). ($\text{b}_2$) The population of state $|E_1\rangle$ (in solid line) and state $|E_2\rangle$ (in dashed line) corresponding to ($\text{b}_1$). The other parameters are set as $\kappa=1$ and $g_A=g_B=0.05$.}
    \label{bt}
\end{figure*}

When the temperatures of the two reservoirs are different, the system is driven into a nonequilibrium environment. We define the temperature difference as $\Delta T=T_A-T_B$ to characterize the degree of nonequilibrium. Under nonequilibrium bosonic environments, the energy output $E_{\text{out}}$ decreases as the average temperature $\bar{T}=(T_A+T_B)/2$ increases, as shown in Fig.~\ref{bt} $(\text{a}_1)$. At low average temperature, $E_{\text{out}}$ decreases with increasing temperature difference $|\Delta T|$, as illustrated by the curve for $\bar{T}=0.5$ in Fig.~\ref{bt} $(\text{a}_1)$. Correspondingly, the population of the ground state remains above 0.9 and is significantly larger than those of the excited states, as shown in Fig.~\ref{bt} $(\text{a}_2)$ (black solid line). As $|\Delta T|$ increases, the ground-state population decreases, mirroring the behavior of $E_{\text{out}}$.

When $\bar{T}=2$, the population of the ground state remains above 0.65, while that of the first excited state stays above 0.2. Unlike the ground-state population, the population of the first excited state increases with $|\Delta T|$. Consequently, the contribution of the first excited state to $E_{\text{out}}$ becomes more pronounced, so that $E_{\text{out}}$ initially increases with $|\Delta T|$. However, when $|\Delta T|$ becomes sufficiently large, the decrease in the ground-state population is no longer compensated mainly by the increase in the population of the first excited state. Instead, higher excited states also acquire non-negligible populations, which leads to a reduction in $E_{\text{out}}$.

As the average temperature further increases to $\bar{T}=5$, the two qubits become separable, and $E_{\text{out}}$ approaches zero with only small variations. Overall, the temperature difference significantly suppresses $E_{\text{out}}$ at low temperatures, whereas at higher temperatures it produces only a slight enhancement.

\begin{figure}[t!]
    \centering
    \includegraphics[width=0.48\textwidth]{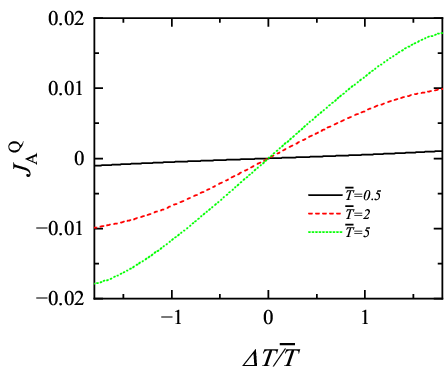}
    \caption{Heat currents $J_A^Q$ between subsystem of Alice and corresponding reservoir. The average temperatures are set as $\bar{T}=0.5$ (black solid line), $\bar{T}=2$ (red dashed line), and $\bar{T}=5$ (green dot line). The other parameters are set as $\varepsilon_A=\varepsilon_B=2$, $\kappa=1$ and $g_A=g_B=0.05$.}
    \label{fbhc}
\end{figure}

The steady state generated in a nonequilibrium environment is accompanied by a heat current. When the reservoir temperatures are equal, the heat current is zero. The detailed balance is preserved and the system is in equilibrium state. When the reservoir temperatures are unequal, heat flows from the higher-temperature reservoir to the lower-temperature reservoir. A nonzero heat current therefore provides a direct signature that the steady state is genuinely nonequilibrium. The heat current \(J_A^Q\) corresponding to the steady states shown in Fig.~\ref{bt}$(\text{a}_1)$ is plotted in Fig.~\ref{fbhc}. Here, a positive value of \(J_A^Q\) means that heat flows from reservoir \(A\) into the system. In the steady state, the two reservoir currents satisfy \(J_A^Q=-J_B^Q\). At equilibrium, \(\Delta T=0\), the heat current vanishes. At nonequilibrium \(\Delta T\neq0\) a finite heat current appears.

To further enhance the nonequilibrium effects, we consider the QET protocol with the initial state prepared from two detuned qubits, namely $\Delta \varepsilon \neq 0$ with $\Delta \varepsilon=\varepsilon_A-\varepsilon_B$, while the average energy splitting is fixed at $\bar{\varepsilon}=(\varepsilon_A+\varepsilon_B)/2$. As illustrated in Fig.~\ref{bt} $(\text{b}_1)$, the energy output $E_{\text{out}}$ initially increases with the detuning $\Delta \varepsilon$. As $\Delta \varepsilon$ increases, the energy level of qubit $A$ rises, allowing more energy to be injected during Alice's measurement. By contrast, the energy level of qubit $B$ limits the maximum extractable energy at Bob's side. Therefore, when $\Delta \varepsilon$ is small, $E_{\text{out}}$ increases with increasing detuning. However, once $E_B$ --- the energy after Bob performs the conditional operation $U_B(u)$ --- becomes sufficiently low, further increasing $\Delta \varepsilon$ reduces the energy output. As shown in Fig.~\ref{bt} $(\text{b}_2)$, the population of $|E_1\rangle$ decreases with $|\Delta \varepsilon|$, while that of $|E_2\rangle$ increases. The influence of temperature on $E_{\text{out}}$ remains significant: increasing temperature enhances thermal excitation in the system and thereby reduces the energy output.

\begin{figure}[t!]
    \centering
    \includegraphics[width=0.48\textwidth]{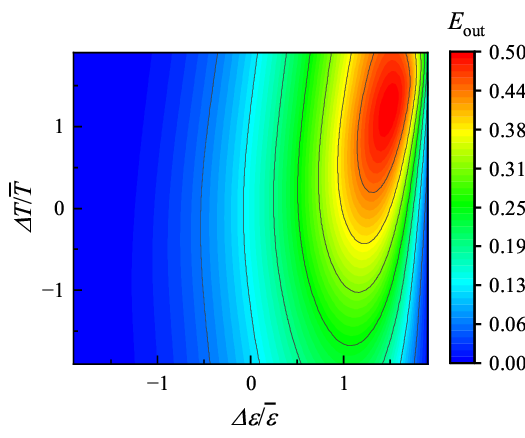}
    \caption{Energy output of steady states of two detuned qubits under the bosonic nonequilibrium environments. The parameters are set as $\bar{\varepsilon}=(\varepsilon_A+\varepsilon_B)/2=2$, $\bar{T}=(T_A+T_B)/2 = 0.5$, $\kappa=1$, and $g_A=g_B=0.05$.}
    \label{bte}
\end{figure}

The combination of energy-level detuning and nonequilibrium environments results in a significant enhancement of $E_{\text{out}}$ at fixed average energy splitting $\bar{\varepsilon}$ and fixed average temperature $\bar{T}$. Specifically, when qubit $A$ has the higher energy level and is coupled to the higher-temperature reservoir, the energy output can be strongly enhanced, as shown in Fig.~\ref{bte}. The influence of the nonequilibrium environment on the detuned two-qubit system is asymmetric. In the region where $E_{\text{out}}$ is enhanced, the ground-state population also increases, because coupling the qubit with the larger energy splitting to the hotter reservoir partially compensates the thermal excitation effect and thus makes the system harder to excite.

\subsection{Nonequilibrium fermionic environments}\label{sec:noneq_fermionic}


\begin{figure*}[t!]
    \centering
    \includegraphics[width=0.9\textwidth]{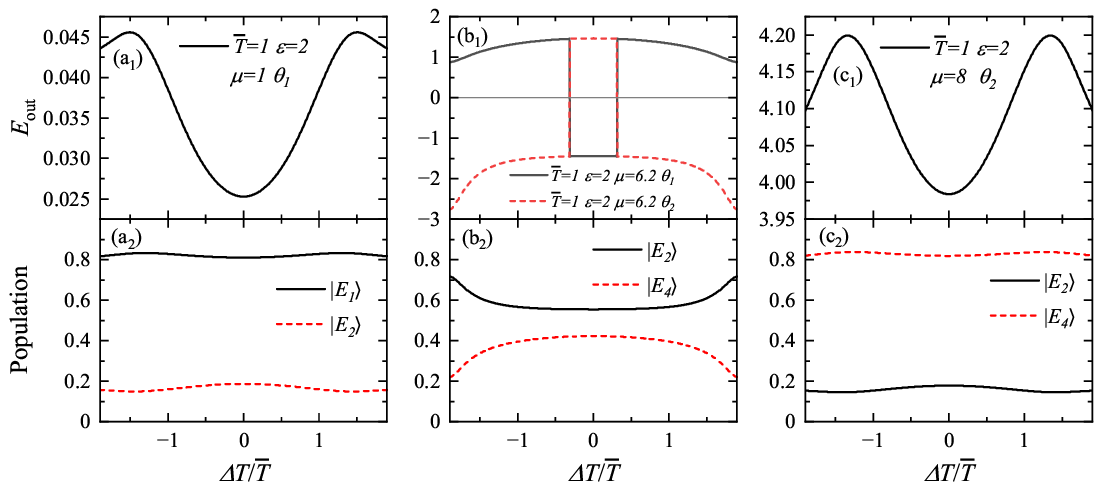}
    \caption{The energy output of steady states under the fermionic reservoirs with nonequilibrium temperatures, and the corresponding population of eigenstates. The chemical potentials are set as ($\text{a}$) $\mu = 1$, ($\text{b}$) $\mu=6.2$, and ($\text{c}$) $\mu=8$. ($\text{a}_2$) The population of state $|E_1\rangle$ (in black solid line) and state $|E_2\rangle$ (in red dashed line) corresponding to ($\text{a}_1$). ($\text{b}_2$) The population of state $|E_2\rangle$ (in black solid line) and state $|E_4\rangle$ (in red dashed line) corresponding to the ($\text{b}_1$). ($\text{c}_2$) The population of state $|E_2\rangle$ (in black solid line) and state $|E_4\rangle$ (in red dashed line) corresponding to the ($\text{c}_1$). The other parameters are set as $\kappa=1$, $\bar T = 1$, and $g_A=g_B=0.05$.}
    \label{ft}
\end{figure*}

When the two qubits are coupled to nonequilibrium fermionic environments, we separately discuss the influence of the temperature difference and the chemical-potential difference on the energy output. When the chemical potential is relatively low, the energy output is determined predominantly by the population of state $|E_1\rangle$, and $E_{\text{out}}$ remains small, as shown in Fig.~\ref{ft} $(\text{a}_1)$. The temperature difference $\Delta T$ suppresses the populations of the excited states and increases the population of $|E_1\rangle$, thereby enhancing $E_{\text{out}}$, as shown in Fig.~\ref{ft} $(\text{a}_2)$. However, for sufficiently large $|\Delta T|$, the population of $|E_1\rangle$ decreases again, which corresponds to a reduction in $E_{\text{out}}$, as shown in Fig.~\ref{ft} $(\text{a}_1)$.

\begin{figure*}[t!]
    \centering
    \includegraphics[width=0.9\textwidth]{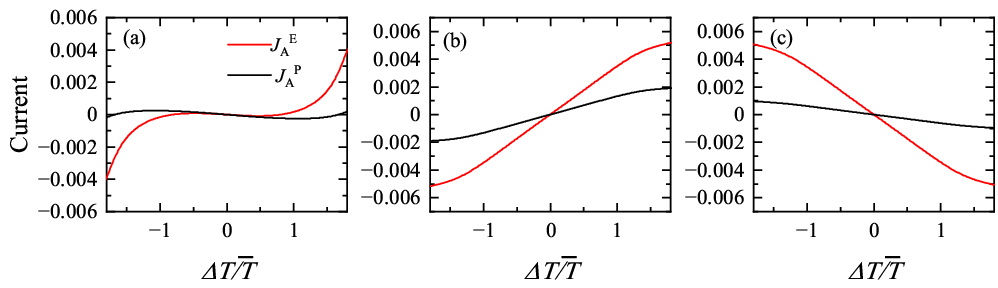}
    \caption{Energy current and particle current between subsystem of Alice and corresponding reservoir. The red lines are heat currents and the black lines are particle currents in subfigures (a) ($\mu=1$), (b) ($\mu=6.2$) and (c) ($\mu=8$). The other parameters are set as $\kappa=1$, $\bar T = 1$, $\varepsilon_A=\varepsilon_B=2$ and $g_A=g_B=0.05$.}
    \label{fhcpct}
\end{figure*}

The steady states generated by nonequilibrium fermionic reservoirs with unequal temperatures are accompanied by both heat and particle currents. The corresponding currents \(J_A^E\) and \(J_A^P\) for the steady states shown in Fig.~\ref{ft} are plotted in Fig.~\ref{fhcpct}. Here, a positive value of \(J_A^E\) or \(J_A^P\) means that energy or particles flow from reservoir \(A\) into the system. At equilibrium, \(\Delta T=0\), both currents vanish. At nonequilibrium \(\Delta T\neq 0\) finite heat and particle currents appear, confirming that the steady states are genuine nonequilibrium steady states. In the steady state, the currents associated with the two reservoirs satisfy \(J_A^E=-J_B^E\) and \(J_A^P=-J_B^P\). The dependence of these currents on \(\Delta T\) also shows that, in fermionic environments, thermal bias not only drives energy transport but can simultaneously induce particle transport.

When the chemical potential is comparatively high, the energy output is governed predominantly by the population of state $|E_4\rangle$, and $E_{\text{out}}$ is therefore significantly enhanced, as illustrated in Fig.~\ref{ft} $(\text{c}_1)$. Increasing the temperature difference $\Delta T$ enhances $E_{\text{out}}$ by facilitating transitions from $|E_2\rangle$ to $|E_4\rangle$. However, in more strongly nonequilibrium cases, i.e., for large $|\Delta T|$, the population of $|E_4\rangle$ decreases, which in turn reduces $E_{\text{out}}$, as shown in Fig.~\ref{ft} $(\text{c}_2)$.

For moderate chemical potential, i.e., when it is comparable to the system energy scale, the energy output exhibits two distinct regimes. As shown in Fig.~\ref{ft} $(\text{b}_1)$, the QET control parameter $\theta$ in the correction operator $U_B(u)$ switches from $\theta_2$ to $\theta_1$ as $|\Delta T|$ increases. In this regime, thermal effects and particle exchange induced by the nonequilibrium fermionic reservoirs act together to promote excitation. In other words, the population of state $|E_4\rangle$ is sufficiently large, so the optimal QET protocol corresponds to $\theta=\theta_2$. As shown in Fig.~\ref{ft} $(\text{b}_2)$, increasing the temperature difference reduces the population of $|E_4\rangle$ because the qubit coupled to the lower-temperature reservoir becomes less likely to be excited, thereby reducing $E_{\text{out}}$. Furthermore, the population lost from $|E_4\rangle$ is transferred mainly to $|E_2\rangle$, rather than to the ground state.

\begin{figure*}[t!]
    \centering
    \includegraphics[width=0.9\textwidth]{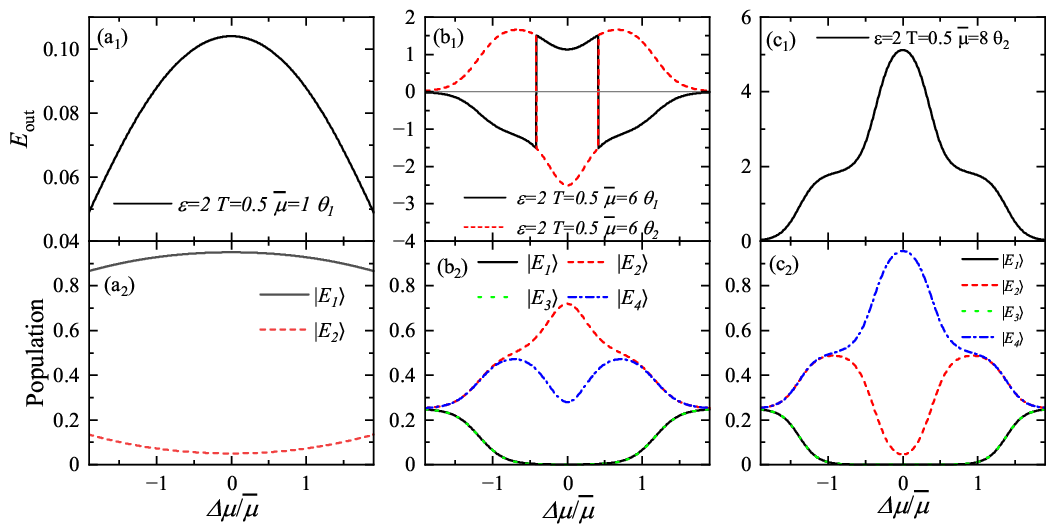}
    \caption{The energy output of steady states under the fermionic reservoirs with nonequilibrium chemical potential, and the corresponding population of eigenstates. The average chemical potentials are set as ($\text{a}$) $\bar{\mu}=1$, ($\text{b}$) $\bar{\mu}=6$, and ($\text{c}$) $\bar{\mu}=8$. ($\text{a}_2$) The population of state $|E_1\rangle$ (in black solid line) and state $|E_2\rangle$ (in red dashed line) corresponding to ($\text{a}_1$). ($\text{b}_2$) The population of state $|E_1\rangle$ (in black solid line), state $|E_2\rangle$ (in red dashed line), state $|E_3\rangle$ (in green dot line), and state $|E_2\rangle$ (in blue dash-dot line) corresponding to ($\text{b}_1$). ($\text{c}_2$) The population of state $|E_1\rangle$ (in black solid line), state $|E_2\rangle$ (in red dashed line), state $|E_3\rangle$ (in green dot line), and state $|E_2\rangle$ (in blue dash-dot line) corresponding to ($\text{c}_1$). The other parameters are set as $\kappa=1$, $T_A=T_B=1$ and $g_A=g_B=0.05$.}
    \label{fmu}
\end{figure*}

When the chemical potentials of the two reservoirs are different, with $\Delta \mu=\mu_A-\mu_B$, the resulting behavior must be analyzed on a case-by-case basis, depending on the value of the average chemical potential. When the average chemical potential $\bar{\mu}=(\mu_A+\mu_B)/2$ is relatively low, the primary effect of $|\Delta \mu|$ is to drive the system from the ground state to the first excited state, as shown in Fig.~\ref{fmu} $(\text{a}_2)$. In this regime, the energy output $E_{\text{out}}$ depends mainly on the ground-state population. Therefore, as $|\Delta \mu|$ increases, the energy output decreases accordingly, as shown in Fig.~\ref{fmu} $(\text{a}_1)$.

As the average chemical potential increases, the population of the highest excited state gradually rises. The energy output $E_{\text{out}}$ is enhanced when $|\Delta \mu|$ is small, but is reduced when the chemical-potential difference becomes large, as shown in Fig.~\ref{fmu} $(\text{b}_1)$. As shown in Fig.~\ref{fmu} $(\text{b}_2)$, the chemical-potential difference $|\Delta \mu|$ can enhance the population of state $|E_4\rangle$ when $|\Delta \mu|\lesssim 0.5$. As $|\Delta \mu|$ increases further, the qubit coupled to the lower-chemical-potential reservoir tends to be de-excited, as reflected in the increasing population of $|E_2\rangle$. When the chemical-potential difference becomes large, the populations of all eigenstates approach each other, resulting in a mutual cancellation of the energy contributions from different eigenstates and hence a vanishing $E_{\text{out}}$.

In the case of a high average chemical potential, as shown in Fig.~\ref{fmu} $(\text{c}_1)$, the variation of $E_{\text{out}}$ is governed mainly by the population of state $|E_4\rangle$. Since the population of $|E_4\rangle$ is predominant in this regime, increasing the chemical-potential difference reduces the population imbalance among the eigenstates, as shown in Fig.~\ref{fmu} $(\text{c}_2)$, thereby decreasing the energy output. Therefore, the chemical-potential difference acts only to suppress the energy output in this regime.

\begin{figure*}[t!]
    \centering
    \includegraphics[width=0.9\textwidth]{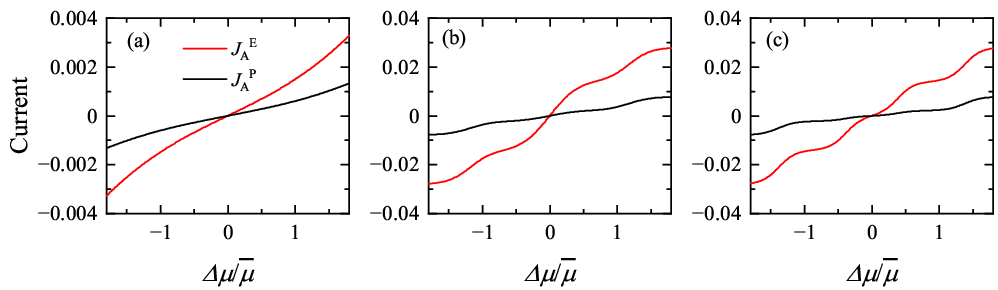}
    \caption{Energy current and particle current between subsystem of Alice and corresponding reservoir. The red lines are heat currents and the black lines are particle currents in subfigures (a) ($\bar{\mu}=1$), (b) ($\bar{\mu}=6$) and (c) ($\bar{\mu}=8$). The other parameters are set as $\kappa=1$, $T_A =T_B =0.5$, $\varepsilon_A=\varepsilon_B=2$ and $g_A=g_B=0.05$.}
    \label{fhpcmu}
\end{figure*}

When the chemical potentials of the two fermionic reservoirs are different, the resulting nonequilibrium steady states are likewise characterized by nonzero heat and particle currents. The corresponding currents \(J_A^E\) and \(J_A^P\) for the steady states shown in Fig.~\ref{fmu} are displayed in Fig.~\ref{fhpcmu}. At \(\Delta\mu=0\), both currents vanish. At nonequilibrium finite chemical potential difference generates particle transport accompanied by a heat current, providing direct evidence that the steady states are genuinely nonequilibrium. In the steady state, one has \(J_A^E=-J_B^E\) and \(J_A^P=-J_B^P\). As the average chemical potential increases, the particle current becomes more pronounced, consistent with the increasingly important role of highly excited states in determining the energy output.

\begin{figure}[t!]
    \centering
    \includegraphics[width=0.48\textwidth]{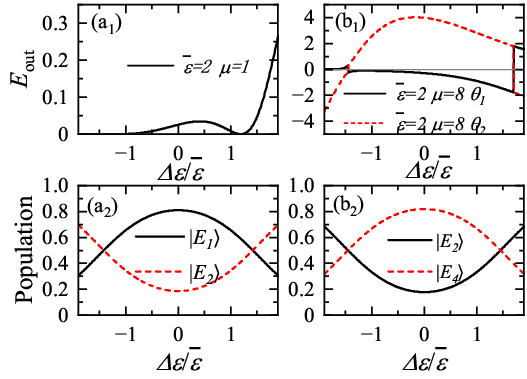}
    \caption{Energy output of steady states with detuned energy levels. The parameters are set as ($\text{a}$) $\mu=1$ and ($\text{b}$) $\mu=8$. ($\text{a}_2$) The population of state $|E_1\rangle$ (in black solid line) and state $|E_2\rangle$ (in red dashed line) corresponding to ($\text{a}_1$). ($\text{b}_2$) The population of state $|E_2\rangle$ (in black solid line) and state $|E_4\rangle$ (in red dashed line) corresponding to ($\text{b}_1$). The other parameters are set as $\kappa=1$,  $\bar{\varepsilon}=2$, $T_A=T_B=1$, and $g_A=g_B=0.05$.}
    \label{fe}
\end{figure}

If we consider two detuned qubits with nonzero $\Delta \varepsilon=\varepsilon_A-\varepsilon_B$, the energy output exhibits an asymmetric dependence on $\Delta \varepsilon$. When the chemical potential is relatively low, the detuning $\Delta \varepsilon$ primarily affects the populations of states $|E_1\rangle$ and $|E_2\rangle$, as shown in Fig.~\ref{fe} $(\text{a}_1)$. For nonzero $\Delta \varepsilon$, the qubit with the lower energy level is more easily excited, leading to an increase in the population of the first excited state $|E_2\rangle$. When $\Delta \varepsilon$ is small, the population of $|E_1\rangle$ remains dominant, and the energy output initially increases with $\Delta \varepsilon$ before decreasing, as shown in Fig.~\ref{fe} $(\text{a}_2)$. The initial increase of $E_{\text{out}}$ arises from the increase in $\varepsilon_A$, whereas the subsequent decrease is associated with the reduction in the population of $|E_1\rangle$. By contrast, when $\Delta \varepsilon$ becomes large, the population of $|E_2\rangle$ becomes dominant. In this regime, the behavior of $E_{\text{out}}$ approaches that of $E_{\text{out}}(|E_2\rangle)$, and is enhanced with increasing $\Delta \varepsilon$ because the population of $|E_2\rangle$ continues to grow.

When the chemical potential is relatively high, the two-qubit system is predominantly populated in excited states, as illustrated in Fig.~\ref{fe} $(\text{b}_1)$. For large detuning $\Delta \varepsilon$, excitation of the individual qubits becomes increasingly difficult. In this case, the population distribution is concentrated mainly in states $|E_2\rangle$ and $|E_4\rangle$, as shown in Fig.~\ref{fe} $(\text{b}_2)$. As $|\Delta \varepsilon|$ increases, the populations of $|E_2\rangle$ and $|E_4\rangle$ change in a compensating manner. When the population of $|E_2\rangle$ becomes dominant, the QET protocol switches the optimal parameter from $\theta_1$ to $\theta_2$, and the curve of $E_{\text{out}}$ correspondingly exhibits a sudden change.

\begin{figure*}[t!]
    \centering
    \includegraphics[width=0.9\textwidth]{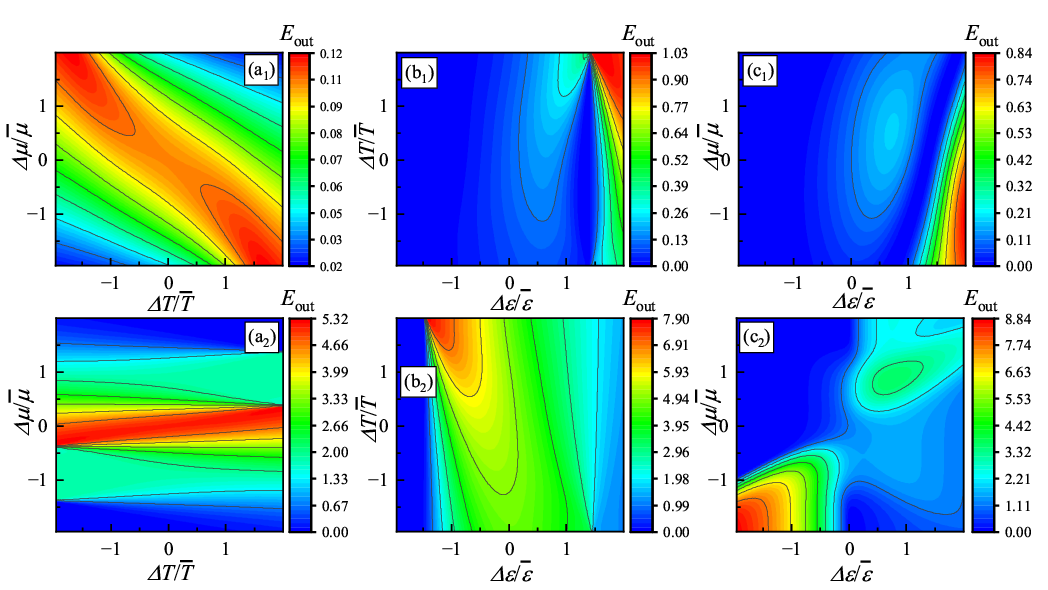}
    \caption{Energy output of steady states with two nonequilibrium parameters. The parameters are set as ($\text{a}_1$): $\bar{T}=0.5, \bar{\mu}=1~\text{and}~\varepsilon_A=\varepsilon_B=2$; ($\text{a}_2$): $\bar{T}=0.5, \bar{\mu}=8~\text{and}~\varepsilon_A=\varepsilon_B=2$; $b_1$: $\bar{\varepsilon}=2, \bar{T}=0.5~\text{and}~\mu_A=\mu_B=1$; $b_2$: $\bar{\varepsilon}=2, \bar{T}=0.5~\text{and}~\mu_A=\mu_B=8$; $c_1$: $\bar{\varepsilon}=2, \bar{\mu}=1~\text{and}~T_A=T_B=0.5$ and $c_2$: $\bar{\varepsilon}=2, \bar{\mu}=6~\text{and}~T_A=T_B=0.5$.The other parameters are set as $\kappa=1$, and $g_A=g_B=0.05$.}
    \label{fd}
\end{figure*}

When both temperature and chemical-potential nonequilibrium are present, and the average chemical potential is low, the variation of $E_{\text{out}}$ correlates mainly with the variation of the ground-state population. The combination of a high (low) temperature reservoir with a low (high) chemical potential increases the ground-state population, thereby enhancing the energy output, as shown in Fig.~\ref{fd} $(\text{a}_1)$. By contrast, for high average chemical potential, the behavior of $E_{\text{out}}$ is governed mainly by the population of $|E_4\rangle$. In this case, the influence of the temperature difference is weak, and $E_{\text{out}}$ is governed primarily by the chemical-potential difference. Accordingly, its distribution resembles that of the case in which only the chemical potential is out of equilibrium, as shown in Fig.~\ref{fd} $(\text{a}_2)$. Overall, $E_{\text{out}}$ is enhanced in specific nonequilibrium regions.


When two detuned qubits are coupled to nonequilibrium environments, the distribution of $E_{\text{out}}$ is no longer centrally symmetric around the equilibrium point. When the reservoir temperatures are different, and the average chemical potential is low, $E_{\text{out}}$ becomes significantly enhanced for large $\Delta T$ as the energy-level difference $\Delta \varepsilon$ increases, owing to the increase in the population of the first excited state, as shown in Fig.~\ref{fd} $(\text{b}_1)$. In addition, coupling the qubit with the higher energy level to the hotter reservoir further enhances the population of state $|E_2\rangle$. In the case of relatively high chemical potential, the magnitude of $E_{\text{out}}$ is governed mainly by the population of the highest excited state. Consequently, $E_{\text{out}}$ can be strongly enhanced in the upper-left region of Fig.~\ref{fd} $(\text{b}_2)$.

When two detuned qubits are coupled to nonequilibrium environments with different chemical potentials, the energy output is enhanced when Bob's qubit is connected to the reservoir with the higher chemical potential.

In the regime of low average chemical potential, the distribution of $E_{\text{out}}$ clearly reflects the combined effects of energy detuning and chemical-potential difference, as shown in Fig.~\ref{fd} $(\text{c}_1)$. By contrast, for high average chemical potential, coupling the qubit with the higher (lower) energy level to the reservoir with the higher (lower) chemical potential yields a larger population of $|E_4\rangle$, which in turn enhances the energy output, as shown in Fig.~\ref{fd} $(\text{c}_2)$. In particular, a larger $\varepsilon_B$ allows more energy to be extracted, thereby leading to a higher energy output in the lower-left quadrant of Fig.~\ref{fd} $(\text{c}_2)$.

\section{Conclusions\label{V}}

In this work, we have investigated the effects of both equilibrium and nonequilibrium parameters on QET in a two-qubit model coupled to two separate environments. By analyzing the behavior of the energy output, we have qualitatively clarified how equilibrium and nonequilibrium environments influence QET performance through the redistribution of energy-eigenstate populations in mixed steady states.

In bosonic reservoirs, we find that when qubit $A$ has a higher energy level and is coupled to the higher-temperature reservoir, the energy input at Alice's side increases, thereby enhancing the energy output. More generally, the influence of nonequilibrium conditions on the energy output is governed mainly by the ground-state population. By analyzing the combined effects of temperature difference and energy-level detuning, we find that the energy output can be significantly enhanced in the region with $\Delta T>0$ and $\Delta \varepsilon>0$, compared with the equilibrium case.

For fermionic reservoirs, the scenario is more complex. At low chemical potential, the energy output is mainly determined by the ground-state population, similar to the bosonic case. However, at high chemical potential, the population of the highest excited state becomes large, and this excited-state population becomes the dominant factor governing the energy output.

For fermionic reservoirs, temperature differences can generally enhance the energy output, whereas chemical-potential differences tend to suppress it. We have also considered the combined effects of temperature and chemical-potential nonequilibrium, as well as the case of two detuned qubits coupled to nonequilibrium environments with either a temperature bias or a chemical-potential bias. In these situations, the energy output $E_{\text{out}}$ can be enhanced in certain parameter regions far from equilibrium. Overall, nonequilibrium conditions in both bosonic and fermionic reservoirs can improve the performance of QET.

Note that the QET protocols for the four eigenstates require different optimal control operations, indicating that the present protocol cannot fully extract energy from a mixture of eigenstates. This suggests that more effective QET strategies may exist for mixed states. Ideally, a more general energy-extraction protocol would operate effectively for all eigenstates and thereby enable larger energy extraction from mixed states. If achieved, such a protocol could also provide valuable insights into the quantum resources underlying QET.

\section{Acknowledgement}
X. K. Yan thanks NSF 12234019 for support. K. Zhang is supported by the National Natural Science Foundation of China under Grant Nos. 12305028 and 12247103, China Postdoctoral Science Foundation under Grant Number 2025M773421, Shaanxi Province Postdoctoral Science Foundation under Grant Number 2025BSHYDZZ017, Scientific Research Program Funded by Education Department of Shaanxi Provincial Government (Program No.24JP186), and the Youth Innovation Team of Shaanxi Universities.

\appendix
\section{Energy extracted by LU operation from 'X' state}

A LP state is a multipartite state from which no energy can be extracted by LU operations. In this work, we implement the QET protocol on mixed states. Since Bob's energy-extraction step in the standard QET protocol is realized by a LU, it is necessary to distinguish whether the mixed states under consideration are LP or not. If an LP state allows energy extraction via the QET protocol, or if a non-LP state yields a larger amount of extractable energy under QET than under LU-only operations, then implementing QET on such mixed states is meaningful at the level of LU operations. In this appendix, we show how much energy can be extracted from $X$-type mixed states using only LU operations.

The only LU operations of Bob can be written as $U=I\otimes U_B$, where $U_B$ is single qubit LU operation 
\begin{align}
   U_B= \begin{pmatrix}
\cos\frac{m}{2} & -e^{il}\sin\frac{m}{2}\\
e^{-il}\sin\frac{m}{2} & \cos\frac{m}{2}.
\end{pmatrix}
\end{align}
The energy extracted by LU operation is 
\begin{align}
    E_{LU}=Tr(H_{AB}\rho_{AB})-Tr(H_{AB}U\rho_{AB}U^\dagger).
\end{align}
If we express the mixed state as a mixture of energy eigenstates, one can get 
\begin{align}
    Tr(H_{AB}\rho_{AB})&=(a+b-c-d)\varepsilon_A+(a-b+c-d)\varepsilon_B \nonumber \\
    &+4\kappa(\delta\cos\epsilon+\chi\cos\nu), \nonumber \\
    Tr(H_{AB}U\rho_{AB}U^\dagger)&=Tr(H_{AB}\rho_{AB})-2\sin^2\frac{m}{2}\Bigl\{(a-b+c-d)\varepsilon_B \nonumber\\
&+4\kappa\cos l\bigl[\chi\cos(\nu-l)+\delta\cos(\epsilon+l)\bigr]\Bigr\}
,
\end{align}
then the extracted energy \( E_b \) is given by 
\begin{align}\label{lueb}
    E_{LU}&=2\sin^2\frac{m}{2}\{(a-b+c-d)\varepsilon_B\nonumber \\
    &+4\kappa\cos l[\chi\cos(\nu-l)+\delta\cos(\epsilon+l)]\}.
\end{align}

To compute the maximum \(E_b\) for a given steady state under arbitrary LU operations, we employ a quantum particle swarm optimization (QPSO) algorithm to search for the optimal value. For each steady state, the parameters are iteratively updated within the domain allowed by LU operations to maximize \(E_b\). The update rule reads
\begin{align}
    x_i(t+1) &= \beta\, p_i^{\mathrm{Best}} + (1-\beta)\, g^{\mathrm{Best}} \nonumber\\
    &\quad \pm \alpha\, \big|m^{\mathrm{Best}}-x_i(t)\big| \ln\!\left(\frac{1}{u}\right),
\end{align}
where \(p_i^{\mathrm{Best}}\), \(g^{\mathrm{Best}}\), and \(m^{\mathrm{Best}}\) denote the individual best, global best, and mean best positions, respectively. The quantum-tunneling-like search is realized through the random variables \(\beta\) and \(u\), while \(\alpha\) is the contraction--expansion coefficient.

\begin{figure*}[htpb]
    \centering
    \includegraphics[width=0.98\textwidth]{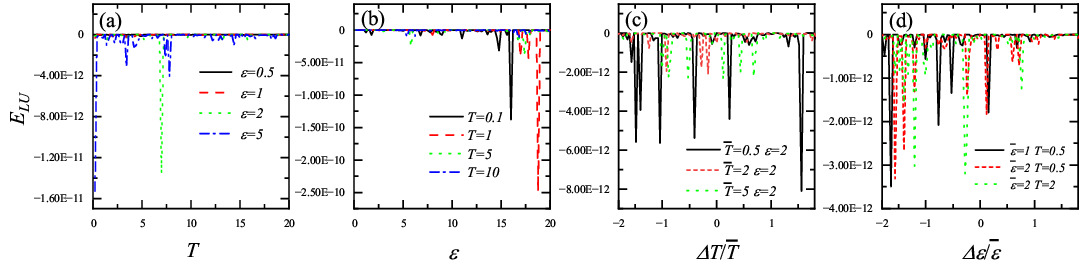}
    \caption{Energy extraction by LU operation of the steady states in boson reservoirs: subfigure (a) and (b) corresponding to Fig. \ref{eqbte}; subfigure (c) and (d) corresponding to Fig. \ref{bt}.}
    \label{ELUB}
\end{figure*}

We calculated the maximal value of \(E_{LU}\) for the steady states in boson reservoirs shown in Figs. \ref{eqbte} (Fig. \ref{ELUB} (a) and (b)) and \ref{bt}  (Fig. \ref{ELUB} (c) and (d)) . Numerically, all these states are LP states, meaning that no energy can be extracted from them by LU operations alone. However, after the QET protocol is applied, they become states from which energy can be extracted via LU operations.

\begin{figure}[htpb]
    \centering
    \includegraphics[width=0.48\textwidth]{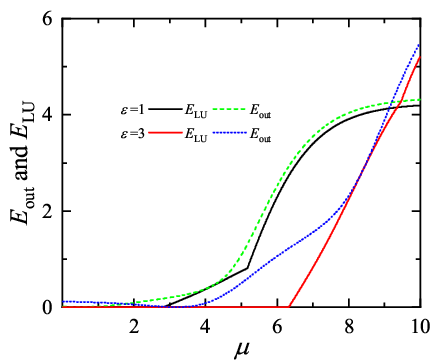}
    \caption{Energy extraction of QET and LU operation of the steady states in equilibrium reservoirs corresponding to Fig. \ref{eqfmu}.}
    \label{AF1}
\end{figure}

Then, we examine the dependence of \(E_b\) on the chemical potential for fermionic reservoirs in thermal equilibrium, as shown in Fig.~\ref{AF1}, which corresponds to the example presented in Fig.~\ref{eqfmu} of the main text. As can be seen, when the chemical potential is relatively low, the system remains in a LP state. As the chemical potential increases, the system becomes locally active, so that energy can be extracted by LU operations. At the same time, the energy extracted via QET is consistently larger than that obtainable using LU operations alone.

\begin{figure*}[htpb]
    \centering
    \includegraphics[width=0.9\textwidth]{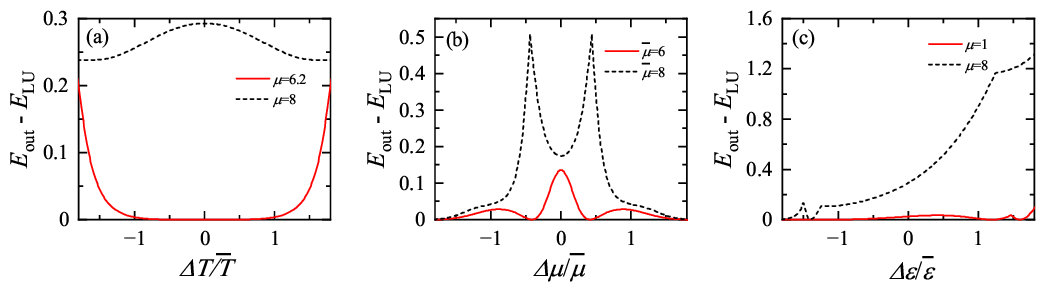}
    \caption{The difference between the energy extraction of QET and LU operation. (a): The steady states in nonequilibrium reservoirs with different temperature in Fig. \ref{ft}. (b): The steady states in nonequilibrium reservoirs with different chemical potential in Fig. \ref{fmu}. (c): The steady states with energy level detuning in Fig. \ref{fe}.}
    \label{AF2}
\end{figure*}

For steady states in high-chemical-potential reservoirs with a temperature difference, a chemical-potential difference, or asymmetric energy levels, we calculate the quantity \(E_{\mathrm{out}}-E_b\), as shown in Fig.~\ref{AF2}, for the corresponding examples in the main text. It is clear that under all these conditions, \(E_{\mathrm{out}}-E_b\) remains positive, which shows that QET always extracts more energy than LU operations alone in these cases.

From the above analysis of \(E_b\), we conclude that, within the model considered here, the standard QET protocol applied to a mixed state can enable energy extraction even when the original system is locally passive. Moreover, when the system is not locally passive, the QET protocol yields more extractable energy than purely local-unitary operations.

\bibliographystyle{elsarticle-num} 
\bibliography{apssamp}

\end{document}